\documentclass[acmtog, authorversion]{acmart}
\acmSubmissionID{1339}

\usepackage{booktabs} 
\usepackage{mathtools,colortbl}
\usepackage{appendix}
\usepackage{bm}
\usepackage{stmaryrd} 

\usepackage[ruled]{algorithm2e} 

\SetAlFnt{\small}
\SetAlCapFnt{\small}
\SetAlCapNameFnt{\small}
\SetAlCapHSkip{0pt}

\acmJournal{TOG}

\copyrightyear{2025} 
\acmYear{2025} 
\setcopyright{cc}
\setcctype{by}
\acmConference[SIGGRAPH Conference Papers '25]{Special Interest Group on Computer Graphics and Interactive Techniques Conference
Conference Papers }{August 10--14, 2025}{Vancouver, BC, Canada}
\acmBooktitle{Special Interest Group on Computer Graphics and Interactive Techniques Conference Conference Papers (SIGGRAPH Conference
Papers '25), August 10--14, 2025, Vancouver, BC, Canada}
\acmDOI{10.1145/3721238.3730754}
\acmISBN{979-8-4007-1540-2/2025/08}

\newcommand{\vect}[1] {\mathbf{#1}}
\newcommand{\cv}{\vect{x}} 
\newcommand{\lv}{\vect{z}} 
\newcommand{\Lv}{\vect{Z}} 
\newcommand{\diro}{\omega_o} 

\newcommand{\netparam}{\theta}

\newcommand{\rev}[1]{\textcolor{black}{#1}}
\newcommand{\revsp}[1]{\textcolor{black}{#1}}

\begin{document}
\title{Neural Importance Sampling of Many Lights}

\author{Pedro Figueiredo}
\orcid{0000-0002-3807-1512}
\affiliation{
\institution{Texas A\&M University}
\country{USA}
}
\email{pedrofigueiredo@tamu.edu}

\author{Qihao He}
\orcid{0009-0000-3011-3135}
\affiliation{
\institution{Texas A\&M University}
\country{USA}
}
\email{phyqh@tamu.edu}

\author{Steve Bako}
\orcid{0000-0002-8090-7918}
\affiliation{
\institution{Aurora Innovation}
\country{USA}
}
\email{sbako@aurora.tech}

\author{Nima Khademi Kalantari}
\orcid{0000-0002-2588-9219}
\affiliation{
\institution{Texas A\&M University}
\country{USA}
}
\email{nimak@tamu.edu}

\renewcommand\shortauthors{Figueiredo et al.}

\begin{abstract}
We propose a neural approach for estimating spatially varying light selection distributions to improve importance sampling in Monte Carlo rendering, particularly for complex scenes with many light sources. Our method uses a neural network to predict the light selection distribution at each shading point based on local information, trained by minimizing the KL-divergence between the learned and target distributions in an online manner. To efficiently manage hundreds or thousands of lights, we integrate our neural approach with light hierarchy techniques, where the network predicts cluster-level distributions and existing methods sample lights within clusters. Additionally, we introduce a residual learning strategy that leverages initial distributions from existing techniques, accelerating convergence during training. Our method achieves superior performance across diverse and challenging scenes.
\end{abstract}

%
%


%
%


\maketitle
\section{Introduction}
\label{sec:intro}

\begin{figure}[t]
\centering
\includegraphics[width=1.0\linewidth]{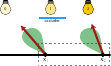}
\vspace{-0.2in}
\caption{We illustrate a scene with three light sources of increasing brightness. Accurate light sampling requires considering factors such as light intensity, the BRDF, and visibility. The light with the highest contribution at each shading point, accounting for all factors, is indicated by red arrows. Sampling based solely on illumination magnitude leads to over-sampling light 2 at shading point $\cv_1$, even though it does not contribute due to the directional constraints of the BRDF. Similarly, methods that consider both light intensity and the BRDF~\cite{Estevez_2018_CGIT, Yuksel_2019_HPG} heavily sample light 1, which is occluded and thus does not contribute to the reflected light. Properly incorporating all three factors reveals that light 0 should be sampled more frequently at $\cv_1$. While learning-based techniques~\cite{Vevoda_2018_TOG, Pantaleoni_2019_arXiv, Wang_2021_TOG} incorporate visibility into their framework, they compute a single distribution for clusters of shading points (indicated by the gray dashed box), failing to capture critical local variations. For instance, light 0 dominates at $\cv_1$, whereas light 2 contributes most at the neighboring point $\cv_2$, underscoring the importance of spatially varying distributions.}
\vspace{-0.2in}
\label{fig:motivation}
\end{figure}

Monte Carlo (MC) rendering is the gold standard for offline rendering and is widely adopted in the film industry, where photorealistic results are critical. With this widespread adoption, the demand for rendering increasingly complex scenes has grown, particularly those with many light sources. Even computing direct illumination in such scenarios presents significant challenges, as efficient rendering requires identifying the light sources that contribute the most at each shading point by accounting for the light intensity, geometry, visibility, and material properties (see Fig.~\ref{fig:motivation}).

A group of techniques~\cite{Paquette_1998_CGF, Fernandez_2002_EG, Walter_2005_TOG, Estevez_2018_CGIT, Liu_2019_CGF, Moreau_2022_HPG, Pantaleoni_2019_arXiv, Vevoda_2018_TOG, Yuksel_2019_HPG, Wang_2021_TOG, Tokuyoshi_2024} address this problem by constructing a light hierarchy (e.g., using a binary tree) where lights are grouped based on criteria such as spatial position, intensity, or other characteristics. These methods then adaptively determine a cut through the hierarchy and define the importance of the clusters along the cut. Clusters are sampled with probabilities proportional to their importance, with either a representative light from the selected cluster being evaluated~\cite{Walter_2005_TOG} or subtrees being traversed to evaluate individual lights~\cite{Estevez_2018_CGIT, Yuksel_2019_HPG}. A key limitation of these approaches is that importance is computed without considering visibility, which can result in frequently sampling occluded lights (see Fig.~\ref{fig:motivation}).

To address this problem, several methods~\cite{Fernandez_2002_EG, Donikian_2006_TVCG, Vevoda_2018_TOG, Pantaleoni_2019_arXiv, Wang_2021_TOG} propose progressively improving the sampling distributions during rendering by incorporating visibility. All these techniques utilize complex spatial data structures to track cluster importance during the online learning process and they mainly differ in their optimization strategies. Despite their ability to account for visibility, these methods optimize a single distribution for groups of shading points in each subspace, ignoring significant variations within the subareas (see Fig.~\ref{fig:motivation}).


In this paper, we address these limitations by estimating a spatially varying light selection distribution using a neural network. Unlike prior approaches that rely on complex spatial data structures to track and optimize cluster importance, our network directly outputs the distribution for each shading point based on its local information (e.g., position and outgoing direction). This eliminates the need for maintaining and querying such data structures, while enabling the estimation of spatially varying distributions across all shading points. To train the network, we minimize the KL-divergence between the learned and target distributions, where the target distribution is approximated using Monte Carlo (MC) sampling.

To efficiently handle scenes with a large number of light sources (hundreds or thousands), we propose combining our neural approach with existing light hierarchy techniques. Specifically, we construct a light hierarchy and define our cut as all the nodes at a certain height of the tree. The neural network predicts the distribution over these clusters, while existing methods are used to sample lights within each cluster. Furthermore, we enhance convergence by utilizing the cluster distribution from existing techniques as an initialization and training the network to estimate the residual. 

Although inspired by neural path guiding methods~\cite{Muller_TOG_2019,Dong_2023,Huang_TOG_2024} that focus on continuous selection of the ray directions, our formulation is specifically designed for discrete light selection. This distinction is critical for handling scenes with a large number of light sources, where finding many important, but sparsely distributed lights, in the directional domain may be challenging.

We demonstrate that our method outperforms existing approaches on diverse and challenging scenes.

In summary, our contributions are as follows:

\begin{itemize}
    \item A neural approach for learning light selection distributions for effective importance sampling.  
    \item A hybrid method that integrates the neural approach with light hierarchy techniques to efficiently manage scenes with large numbers of light sources.  
    \item A residual learning strategy to reduce the startup cost and improve convergence during training.  
\end{itemize}

\section{Related Work}
\label{sec:relatedwork}

\paragraph{Many Light Rendering}

Early approaches~\cite{Kok_1994_PRCG, Ward_1994_PRCG, Shirley_1996_TOG} address scenes with many lights by distinguishing between important and unimportant light sources, thereby reducing the number of shadow rays traced. Several methods propose clustering lights by sampling the entries of the light transport matrix~\cite{Hasan_2007_TOG, Hasan_2008_CGF, Ou_2011_TOG, Huo_2015_TOG}, assuming that it is of low-rank.

A more widely adopted approach to tackle this problem involves building a light hierarchy to form clusters~\cite{Paquette_1998_CGF, Fernandez_2002_EG}. Specifically, Walter et al.~\shortcite{Walter_2005_TOG} define clusters by finding a cut through the light tree and sampling representative lights at each cluster. Subsequent works~\cite{Estevez_2018_CGIT, Liu_2019_CGF, Moreau_2022_HPG, Yuksel_2019_HPG, Tokuyoshi_2024} make the sampling process unbiased by traversing the tree stochastically and sampling individual light sources. However, a significant limitation of these techniques is their failure to account for visibility during sampling.

\paragraph*{Learning-Based Light Sampling}
A large number of recent methods incorporate visibility into light sampling in a data-driven manner, either through preprocessing~\cite{Georgiev_2012_CGF, Wu_2013_TVCG} or online updates~\cite{Fernandez_2002_EG, Donikian_2006_TVCG, Vevoda_2018_TOG, Pantaleoni_2019_arXiv, Wang_2021_TOG}. These approaches typically rely on spatial data structures to track distributions, learning a single distribution for a group of shading points within a subregion. However, this approach fails to capture fine-grained variations in the distribution. We address this limitation by learning spatially varying distributions using a neural network.

\paragraph*{Resampled Importance Sampling}
Resampled importance sampling (RIS)~\cite{Talbot_2005_EGSR} provides a framework for rendering methods to importance sample from complex distributions. It achieves this by first drawing samples from a simple proposal distribution, then selecting a subset of these samples based on weights derived from a complex, unnormalized distribution that accounts for factors such as material, geometry, and visibility at each shading point. Since its introduction, resampling techniques have been extended to real-time rendering, most notably in spatiotemporal reservoir resampling (ReSTIR)~\cite{Bitterli_2020_TOG} and its variants for global illumination, including ReSTIR-GI~\cite{Ouyang_2021_CGF} and ReSTIR-PT~\cite{Lin_2022_TOG}. These methods demonstrate exceptional real-time performance in complex, many-light scenarios. Many-light sampling approaches, including ours, are complementary to these techniques, as they can be used to improve the candidate distributions for resampled importance sampling frameworks like RIS and ReSTIR, ultimately enhancing their results.

\paragraph*{Path Guiding}
Path guiding techniques perform importance sampling over the directional domain. Some methods rely on data structures, such as binary and quad trees~\cite{Muller_CGF_2017, Vorba_2019, Rath_TOG_2020, Zhu_TOG_2021}, to store fitted distributions. More recent approaches utilize neural networks to fit target distributions, either offline~\cite{Bako_CGF_2019, Huo_TOG_2020, Zhu_TOG_2021, Zhu_TOG_2021_B} or online~\cite{Muller_TOG_2019, Dong_2023, Huang_TOG_2024, Lu_2024_TOG}. However, fitting accurate distributions for scenes with many sparsely distributed lights remains challenging, limiting the effectiveness of these techniques in such scenarios. Nonetheless, path guiding methods are complementary to light sampling approaches, as their combination with multiple importance sampling (MIS) is often required to handle diverse scenes effectively. While our approach draws inspiration from neural path guiding, we diverge by focusing on a novel formulation for the discrete domain, rather than learning a continuous distribution over the 2D directional domain.

\section{Neural Light Selection Importance Sampling}
\label{sec:method}

We aim to estimate the reflected light, $L_o(\cv, \diro)$, in direction $\diro$ at point $\cv$, due to direct lighting. This is defined as the sum of contributions from all the light sources: 

\vspace{-0.1in}
\begin{equation}
\label{eq:direct_light}
    L_o(\cv, \diro) = \sum_{y = 1}^M L_y(\cv, \diro) = \sum_{y=1}^M \int_{A_y} F(\cv, \lv, \diro) \ dA_y(\lv)
\end{equation}
\vspace{-0.1in}

\noindent where $M$ is the total number of light sources. Moreover, the contribution of $y^{\text{th}}$ light source is obtained by integrating $F$ over its area $A_y$. The integrand, $F$, models the interaction of light emitted from $\lv$, a point on the light source, with the surface and is defined as: 

\vspace{-0.05in}
\begin{equation}
    F(\cv, \lv, \diro) = L_i (\cv, \lv) \ f_s(\cv, \diro, \lv) \ G(\cv, \lv),
\end{equation}
\vspace{-0.05in}

\noindent where $L_i (\cv, \lv)$ represents the radiance emitted from $\lv$ toward $\cv$, $f_s(\cv, \diro, \lv)$ refers to the bidirectional reflectance distribution function (BRDF), and $G(\cv, \lv)$ accounts for geometry and visibility terms.

To facilitate efficient computation of the summation in Eq.~\ref{eq:direct_light}, we express it as an expectation over $Y\sim p(y)$ as follows:

\vspace{-0.0in}
\begin{equation}
    L_o(\cv, \diro) = \mathbb{E}\left [ \frac{L_Y(\cv, \diro)}{p(Y)}\right ]
\end{equation}
\vspace{-0.0in}

\noindent where $p(y)$ is the probability mass function (PMF) and defines the probability of selecting the $y^\text{th}$ light at point $\cv$ and for outgoing direction $\diro$. For brevity, we omit the explicit conditioning on $\cv$ and $\diro$ in the notation of the distribution, though these dependencies remain implied. Note that this equality is valid as long as $p(y)$ is nonzero wherever the light contribution $L_y(\cv, \diro)$ is nonzero.

This expectation can be approximated using $N$ independent samples of light indices, $Y_j \in \{1, \dots, M\}$ and $Y_j \sim p(y)$, with the following Monte Carlo (MC) estimator:

\vspace{-0.05in}
\begin{equation}
\label{eq:direct_light_MC}
    \langle L_o(\cv, \diro) \rangle = \frac{1}{N} \sum_{j = 1}^N \frac{L_{Y_j}(\cv, \diro)}{p(Y_j)}.
\end{equation}
\vspace{-0.05in}

\noindent The choice of $p(y)$ has a substantial impact on the variance of this estimator. To reduce the variance, the light selection PMF, $p(y)$, should closely match the distribution of the numerator. To achieve this, we propose to model $p(y)$ using a neural network, $p_\netparam(y)$, parameterized by $\netparam$, and train it in an online manner.

Note that computing $L_y(\cv, \diro)$ requires evaluating the integral in Eq.~\ref{eq:direct_light} and is also performed through MC integration by randomly sampling a point $\lv$ on the light source. As in prior work on light sampling~\cite{Wang_2021_TOG,Yuksel_2019_HPG,Estevez_2018_CGIT}, we use standard techniques~\cite{Shirley_1996_TOG,Hart_2020_CGF} to perform this.

In the subsequent sections, we detail our optimization strategy for training $p_\netparam(y)$ and describe our efficient network architecture.

\subsection{Optimizing the Light Selection Network}
\label{sec:optimize_light_selection}

Our goal is to train our light selection network so that our learned PMF $p_\netparam(y)$ closely matches the target distribution $q(y)$, the details of which will be discussed later. To this end, we propose to minimize the KL-divergence between the target and learned PMFs as follows: 

\vspace{-0.05in}
\begin{equation}
    D_{\text{KL}}(q, p_\netparam) = \sum_{y = 1}^M q(y) \log\frac{q(y)}{p_\netparam(y)}.
\end{equation}
\vspace{-0.05in}

To minimize this loss, we need to take its derivative with respect to the network parameters $\netparam$ as follows:

\vspace{-0.05in}
\begin{equation}
    \nabla_\netparam D_{\text{KL}}(q, p_\netparam) = - \sum_{y = 1}^M q(y) \nabla_\netparam \log\ p_\netparam(y),
\end{equation}
\vspace{-0.05in}

\noindent where the gradient of $q(y)$ is omitted as the target PMF is independent of $\netparam$. Expressing this gradient as an expectation over $Y\sim p_\netparam(y)$:

\vspace{-0.05in}
\begin{equation}
    \nabla_\netparam D_{\text{KL}}(q, p_\netparam) = - \ \mathbb{E} \left [ \frac{q(Y)}{p_\netparam(Y)} \nabla_\netparam \log\ p_\netparam(Y) \right ],
\end{equation}
\vspace{-0.05in}

\noindent allows us to effectively approximate it using an MC estimator:

\vspace{-0.0in}
\begin{equation}
    \langle \nabla_\netparam D_{\text{KL}}(q, p_\netparam) \rangle = - \frac{1}{N} \sum_{j = 1}^N \left [ \frac{q(Y_j)}{p_\netparam(Y_j)} \nabla_\netparam \log\ p_\netparam(Y_j) \right ],
\end{equation}
\vspace{-0.0in}

\noindent where the samples $Y_j$ are drawn from the learned PMF $p_\netparam(y)$.
We now explain our process of obtaining the target distribution. As discussed, to reduce the variance of the MC estimator in Eq.~\ref{eq:direct_light_MC}, we would like our learned distribution to match the light contributions. Therefore, we define the target distribution as follows:

\vspace{-0.05in}
\begin{equation}
    q(y|\cv, \diro) = L_y(\cv, \diro) L_o(\cv, \diro)^{-1}.
\end{equation}
\vspace{-0.05in}

\noindent Here, $L_y(\cv, \diro)$ represents the reflected radiance at point $\cv$ in direction $\diro$ caused by a single light source. Similarly, $L_o(\cv, \diro)$ denotes the radiance due to all light sources combined (see Eq.~\ref{eq:direct_light}). Note that the second term in the above equation ensures that $q$ is a valid PMF (summing to one).

Obtaining $L_y(\cv, \diro)$ requires evaluating the integral in Eq.\ref{eq:direct_light}, which we approximate using Monte Carlo (MC) integration with point sampling. Specifically, we sample the integrand, $F(\cv, \lv, \diro)$, at randomly selected points $\lv$ on the $y^{\text{th}}$ light. The second term, $L_o(\cv, \diro)$, is unknown because it represents the quantity we aim to estimate. However, since our optimization process relies on Adam~
\cite{Kingma_adam_2015}, which updates parameters based on the ratio of the current gradient to its historical average, this term can safely be omitted~\cite{Muller_TOG_2019,Dong_2023}.

Combining all the components, the final MC estimate of the gradient can be written as:

\vspace{-0.05in}
\begin{equation}
\label{eq:gradient_light_selection}
    \langle \nabla_\netparam D_{\text{KL}}(q, p_\netparam) \rangle = - \frac{1}{N} \sum_{j = 1}^N  \biggl [ \frac{F(\cv, \Lv_j, \diro)}{p(\Lv_j|Y_j)p_\netparam(Y_j)} 
     \nabla_\netparam \log p_\netparam(Y_j) \biggr ].
\end{equation}
\vspace{-0.05in}

\noindent Here, $p(\lv|y)$ is the probability of selecting point $\lv$ on the $y^\text{th}$ light source, which is evaluated using existing methods~\cite{Shirley_1996_TOG,Hart_2020_CGF} and is not learned in our method.

This formulation shows that optimizing the loss is equivalent to minimizing the negative log likelihood weighted by the MC estimates of the integrand $F$. Therefore, the loss can be efficiently optimized during path tracing in an online manner.

\begin{figure*}[t]
\centering
\includegraphics[width=1.0\linewidth]{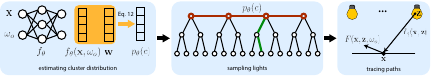}
\vspace{-0.2in}
\caption{We present an overview of our technique. (left) The network uses local information at the shading point to estimate cluster probabilities in a residual manner. The baseline cluster distribution, $\mathbf{w}$, is precomputed using existing approaches and remains fixed throughout the process. (middle) A light is sampled in two stages: first, a cluster (one of the red nodes) is selected based on the estimated cluster distribution, and then the tree is stochastically traversed (green lines) to choose a specific light within the cluster. (right) Points are subsequently sampled within the selected light source, and paths are traced. These samples are used to train the network and refine the estimated cluster distribution. This entire process is repeated iteratively until the desired number of samples is traced. Since the network estimates the cluster PMF in a residual manner, it can guide sampling from the beginning of the rendering process.}
\vspace{-0.05in}
\label{fig:overview}
\end{figure*}

\subsection{Network Architecture}

Our network, implemented in \rev{tiny-cuda-nn}~\cite{tiny-cuda-nn}, is composed of 3 hidden layers of 64 neurons and ReLU intermediary activation. We encode the inputs using ~\cite{Dong_2023}'s learnable dense grid encoding for the intersection position, and spherical harmonics with degree 4 for normalized outgoing ray direction $\diro$ so we can model the product of radiance and BSDF. We additionally provide normal as the input to the network as done in similar learning tasks~\cite{Dong_2023, Muller_TOG_2019}, and encode them through one-blob encoding \cite{Muller_TOG_2019} using 32 bins.

\section{Neural Light Cluster Sampling}
\label{sec:lightclustersampling}

The network described in the previous section directly estimates the probability of selecting each light source. While this approach is feasible when the number of light sources $M$ in the scene is relatively small (e.g., order of tens), directly estimating the PMF for a large number of light sources (e.g., hundreds or thousands) requires a high-capacity network, which significantly increases computational complexity and makes the approach impractical.

Inspired by existing methods~\cite{Estevez_2018_CGIT,Yuksel_2019_HPG}, we address this issue by adopting a clustering strategy. Specifically, we group the lights into a smaller set of $S$ non-overlapping clusters, $\mathcal{L}_c$, with $c \in {1, \dots, S}$. This allows us to decompose the light selection PMF $p(y)$ into two distributions: the probability of selecting cluster $c$, $p(c)$, and the conditional probability of selecting the $y^\text{th}$ light within cluster $c$, $p(y|c)$. We use the network to model the cluster selection PMF $p_\netparam(c)$, while the light selection within each cluster is efficiently handled using existing hierarchy techniques~\cite{Estevez_2018_CGIT,Yuksel_2019_HPG}.

A key advantage of this clustering approach is that it allows us to leverage existing methods to construct the light hierarchy, which also provides initial cluster probabilities. Building on this, we propose a residual learning strategy, where the network learns corrections to these initial probabilities rather than predicting them from scratch, significantly improving training convergence.

The overview of our approach is illustrated in Fig.~\ref{fig:overview}. In the following sections, we describe the light hierarchy and clustering, the network optimization process, and our residual learning strategy.

\subsection{Light Hierarchy and Cluster Construction}

\begin{figure}[t]
\centering
\includegraphics[width=1.0\linewidth]{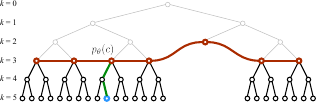}
\vspace{-0.2in}
\caption{Given a light hierarchy constructed using an existing method~\cite{Estevez_2018_CGIT,Yuksel_2019_HPG}, we select nodes at a specific level ($k = 3$ in this example) as light clusters, with the network estimating their probabilities $p_\netparam(c)$. The probability of selecting an individual light (e.g., the blue leaf node) is computed as the product of the cluster probability and the conditional probability of the light given the cluster, $p(y|c)$. The conditional probability is obtained by multiplying the probabilities of the child nodes along the tree traversal path (green lines). Note that the grayed-out nodes are used solely for defining the global clusters and are not involved in sampling or evaluating the PMF.}
\vspace{-0.15in}
\label{fig:light_hierarchy}
\end{figure}

Given a scene, we first construct a light hierarchy using an existing technique~\cite{Estevez_2018_CGIT, Yuksel_2019_HPG}. The nodes at a specified level $k$ of the tree are selected as light clusters (global; fixed for the entire scene), and our network estimates their probabilities. As shown in Fig.~\ref{fig:light_hierarchy}, the tree is typically unbalanced in practical scenarios, so the number of internal nodes at level $k$ may not always equal $2^k$. In such cases, we include leaf nodes from higher levels as part of the clusters, allowing the network to directly estimate the probability of a leaf node that corresponds to a single light source. For instance, in the example shown in Fig.~\ref{fig:light_hierarchy}, the network outputs probabilities for the 7 cluster nodes (highlighted in red).

To sample a light source, we first select a cluster based on the cluster selection PMF $p_\netparam(c)$. Next, we traverse the tree from the selected cluster to a leaf node, selecting one child node at each step based on their PMFs. These PMFs are computed using the same method used to construct the light hierarchy~\cite{Estevez_2018_CGIT, Yuksel_2019_HPG}. The conditional light selection PMF $p(y|c)$ is then determined by multiplying the probabilities of all child nodes along the traversal path, as illustrated by the green lines in Fig.~\ref{fig:light_hierarchy}.

\subsection{Optimization}

We optimize the cluster selection network using the formulation described in Sec.~\ref{sec:optimize_light_selection}. Consequently, the gradient of the KL-divergence loss can be approximated in a manner similar to Eq.~\ref{eq:gradient_light_selection}. The key difference is that the light selection PMF is now expressed as the product of the cluster selection PMF, estimated by the network, and the light selection PMF within a cluster, computed using existing methods. Substituting $p_\netparam(y)$ with $p(y|c)p_\netparam(c)$ in Eq.~\ref{eq:gradient_light_selection}, we derive the approximated gradient:

\vspace{-0.1in}
\begin{align}
    \langle \nabla_\netparam D_{\text{KL}}(q, p_\netparam) \rangle = - \frac{1}{N} \sum_{j = 1}^N  \biggl [ \frac{F(\cv, \Lv_j, \diro)}{p(\Lv_j|Y_j)p(Y_j|C_j)p_\netparam(C_j)} \biggr. \nonumber \\
    \biggl. \nabla_\netparam \log \Bigl ( p(Y_j|C_j)p_\netparam(C_j) \Bigr ) \biggr ]. \nonumber
\end{align}
\vspace{-0.1in}

Since $p(y|c)$ is independent of $\netparam$, we can simplify the $\log$ term as:

\vspace{-0.1in}
\begin{align}
    \langle \nabla_\netparam D_{\text{KL}}(q, p_\netparam) \rangle = - \frac{1}{N} \sum_{j = 1}^N  \biggl [ \frac{F(\cv, \Lv_j, \diro)}{p(\Lv_j|Y_j)p(Y_j|C_j)p_\netparam(C_j)} \nonumber \\
    \nabla_\netparam \log p_\netparam(C_j) \biggr ]. 
\end{align}
\vspace{-0.1in}

By optimizing this objective, the network increases the likelihood of clusters that correspond to light samples with high contributions, while reducing the likelihood of clusters associated with light samples with low contributions.

\subsection{Residual Learning}

As discussed, training is performed in an online manner during the rendering process. We use the randomly initialized network to generate samples that are used as training data to update the network. This process is iteratively continued until a desired number of samples are traced. However, this process has two major issues: \textbf{(1)} the network requires a large number of iterations to converge to a meaningful PMF due to its random initialization, and \textbf{(2)} during this phase, the generated samples exhibit high variance, increasing the noise in final rendering and hindering effective training.

To address these issues, we observe that existing methods~\cite{Estevez_2018_CGIT, Yuksel_2019_HPG} estimate cluster probabilities using carefully designed rules, which are significantly better than the initial PMF predicted by the network. Therefore, we propose to use these cluster probabilities as a baseline, with the network predicting the residual. When combining the network's estimate with the baseline, two key requirements must be met: \textbf{(1)} all cluster probabilities must be positive, and \textbf{(2)} their sum must equal one.

We satisfy the first requirement by performing the combination in the $\log$ domain, ensuring that all values remain positive after applying the inverse $\log$ operation, while the second requirement is met by normalization. Specifically, given the importance weights (where probabilities are normalized importance weights) for all nodes in a cluster, $\vect{w} = {w_1, \dots, w_S}$, we compute the final cluster PMF by combining the network's output with these importance weights as follows:

\vspace{-0.05in}
\begin{equation}
    p_\netparam(c) =  \frac{e^{\left[ \log (w_c) + f_\netparam(\cv, \diro)[c] \right]}}{\sum_{s=1}^S e^{\left[ \log (w_s) + f_\netparam(\cv, \diro)[c] \right]}},
\end{equation}
\vspace{-0.0in}

\noindent where $f_\netparam(\cv, \diro)[c]$ is the estimated residual for the $c^\text{th}$ cluster.

This approach addresses the aforementioned issues effectively. Initially, the cluster probabilities are close to those estimated by existing methods because the randomly initialized network produces outputs near zero. This ensures that the generated samples at the start are comparable to those from the baseline approach. As training progresses, the network refines its predictions, resulting in cluster PMFs that consistently improve upon the baseline.

\section{Results}
\label{sec:results}
In this section, we present visual and numerical comparisons to demonstrate the effectiveness of our approach. Specifically, we compare our method against ATS~\cite{Estevez_2018_CGIT}, SLCRT~\cite{lin2020real}, \rev{ReSTIR~\cite{Bitterli_2020_TOG},} and VARL~\cite{Wang_2021_TOG}. The ATS implementation~\cite{Estevez_2018_CGIT} is available in PBRT v4, while we have implemented the other \rev{three} approaches. \rev{ReSTIR is implemented with $M=32$ samples with $K=3$ spatial reuse samples on a radius of 30 pixels. For VARL, which is designed for CPU, we convert their sequential update to a batch update to make it compatible with GPU}. Our \rev{VARL} implementation is significantly faster than the original version and produces better results (see supplementary material).

\subsection{\rev{Comparisons}}
\label{sec:comparisons}

We perform \rev{equal-time and} equal-sample comparisons on eight scenes that contain complex lighting. \rev{The number of lights per scene is: \textsc{Bathroom} (4.7k), \textsc{Bedroom} (7.9k), \textsc{Bistro} (21.1k), \textsc{LivingRoom} (64), \textsc{Staircase} (6.7k), \textsc{Staircase2} (2.1k) and \textsc{ZeroDay} (8.9k).} For \textsc{San Miguel}, we trace \rev{139k} virtual point lights (VPLs). For all the results, we compare direct illumination at first non-specular bounce. We perform all comparisons and ablations on an Intel i9-14900K CPU and an NVIDIA RTX4090 GPU. Here, we report quantitative results using \reflectbox{F}LIP~\cite{Andersson_2020}. Evaluation using additional metrics can be found in the supplementary. In all comparisons, we use ATS~\cite{Estevez_2018_CGIT} to construct the light hierarchy, and \rev{we estimate node probabilities using our network at level $k=6$. For optimization, we use the Adam~\cite{Kingma_adam_2015} optimizer with default parameters and a fixed learning rate of $3\times10^{-2}$, and choose a training budget of 15\% for our method}.

\begin{table}[t]
\setlength\tabcolsep{3pt} 
\centering
\caption{\rev{Equal-time comparisons measured in \reflectbox{F}LIP~\cite{Andersson_2020} with number of samples per pixel for eight representative scenes. We compare our approach against ATS~\cite{Estevez_2018_CGIT}, SLCRT~\cite{lin2020real}, ReSTIR~\cite{Bitterli_2020_TOG}, and VARL~\cite{Wang_2021_TOG}. We color code the \colorbox{red!40}{first}, \colorbox{orange!40}{second}, and \colorbox{yellow!40}{third} lowest numbers.}}
\vspace{-0.05in}
\footnotesize
\label{tab:equal_time}
\begin{tabular}{lcccccccccc}
\toprule
 & \multicolumn{2}{c}{ATS} & \multicolumn{2}{c}{SLCRT} & \multicolumn{2}{c}{ReSTIR} & \multicolumn{2}{c}{VARL} & \multicolumn{2}{c}{Ours} \\ 
\midrule
\textsc{Bathroom} & 0.1659 & 212 & \cellcolor{yellow!40}0.1413 & 220 & 0.1744  & 135 & \cellcolor{orange!40}0.1346  & 139 & \cellcolor{red!40}0.0977 & 175\\ 
\midrule
\textsc{Bedroom} & 0.3129 & 192 &   \cellcolor{yellow!40}0.2442  & 196 & 0.2796  & 123 & \cellcolor{orange!40}0.1960 & 123 &  \cellcolor{red!40}0.0910  & 157 \\ 
\midrule
\textsc{Bistro} & \cellcolor{orange!40}0.1548 & 163 &  0.2656 & 165 &  0.2837  & 74 & \cellcolor{yellow!40}0.2281 & 60 &  \cellcolor{red!40}0.1126 & 168 \\ 
\midrule
\textsc{Living Room} & 0.1247 & 194 &  \cellcolor{orange!40}0.0977  & 194 &  0.1314  & 131 &\cellcolor{yellow!40}0.1009  & 131 &  \cellcolor{red!40}0.0671  & 146 \\ 
\midrule
\textsc{San Miguel} & \cellcolor{orange!40}0.2253 & 159 &  0.3015 & 158 &  0.2817  & 110 &\cellcolor{yellow!40}0.2472 & 107 & \cellcolor{red!40}0.1713 & 164 \\ 
\midrule
\textsc{Staircase} & \cellcolor{orange!40}0.0763  & 198 &  \cellcolor{yellow!40}0.0817 & 203 & 0.1059  & 129 & 0.0946  & 138 &  \cellcolor{red!40}0.0529 & 160 \\ 
\midrule
\textsc{Staircase2} & 0.1020  & 233 &  \cellcolor{yellow!40}0.0925 & 233 &  0.1132  & 159 &\cellcolor{orange!40}0.0854  & 149 & \cellcolor{red!40}0.0646 & 191 \\ 
\midrule
\textsc{Zero Day} & 0.1647  & 222 &  \cellcolor{yellow!40}0.1455 & 223 &  0.1682  & 159 & \cellcolor{orange!40}0.1338  & 161 & \cellcolor{red!40}0.1117 & 201 \\ 
\bottomrule
\end{tabular}%
\vspace{-0.1in}
\end{table}

\begin{table}[t]
\setlength\tabcolsep{3pt} 
\centering
\caption{Equal-sample comparisons measured in \reflectbox{F}LIP~\cite{Andersson_2020} \rev{with runtime in seconds}. All the scenes are rendered using each approach for 128 samples per pixel.}
\vspace{-0.05in}
\footnotesize
\label{tab:equal_spp}
\begin{tabular}{lcccccccccc}
\toprule
 & \multicolumn{2}{c}{ATS} & \multicolumn{2}{c}{SLCRT} & \multicolumn{2}{c}{ReSTIR} & \multicolumn{2}{c}{VARL} & \multicolumn{2}{c}{Ours} \\ 
\midrule
\textsc{Bathroom} & 0.2059 & 2.8 & 0.1791 & 2.9 & \cellcolor{yellow!40}0.1780  & 4.7 & \cellcolor{orange!40}{0.1399}  & 4.3 & \cellcolor{red!40}0.1090 & 3.9\\ 
\midrule
\textsc{Bedroom} & 0.3780 & 3.2 &  0.2962  & 3.2 &  \cellcolor{yellow!40}0.2757  & 5.2 & \cellcolor{orange!40}0.1913 & 4.8 &  \cellcolor{red!40}0.0994  & 4.4 \\ 
\midrule
\textsc{Bistro} & \cellcolor{yellow!40}0.1704 & 15.9 &  0.2944 & 15.3 & 0.2316  & 36.8 &  \cellcolor{orange!40}0.1658 & 44.3 &  \cellcolor{red!40}0.1180 & 19.6 \\ 
\midrule
\textsc{Living Room} & 0.1488 & 3.2 &  \cellcolor{yellow!40}0.1162  & 3.3 &  0.1387  & 5.2 & \cellcolor{orange!40}0.1023  & 4.6 &  \cellcolor{red!40}0.0689  & 4.8 \\ 
\midrule
\textsc{San Miguel} & \cellcolor{yellow!40}0.2454 & 7.9 &  0.3233 & 8.1 & 0.2675  & 12.4 & \cellcolor{orange!40}0.2308 & 11.7 & \cellcolor{red!40}0.1842 & 8.6 \\ 
\midrule
\textsc{Staircase} & \cellcolor{orange!40}0.0923  & 3.3 &  0.1002  & 3.2 & 0.1060  & 4.9 & \cellcolor{yellow!40}0.0969  & 4.7 &  \cellcolor{red!40}0.0553 & 4.5 \\ 
\midrule
\textsc{Staircase2} & 0.1246  & 2.7 &  \cellcolor{yellow!40}0.1128 & 2.8 &  0.1206 & 4.0 & \cellcolor{orange!40}0.0913  & 4.1 & \cellcolor{red!40}0.0721 & 3.8 \\ 
\midrule
\textsc{Zero Day} & 0.1929  & 5.1 &  \cellcolor{yellow!40}0.1699  & 5.0 &  0.1771  & 8.1 & \cellcolor{orange!40}0.1465  & 7.0 & \cellcolor{red!40}0.1224 & 7.0 \\ 
\bottomrule
\end{tabular}%
\vspace{-0.05in}
\end{table}

\begin{figure*}
\includegraphics[width=\textwidth]{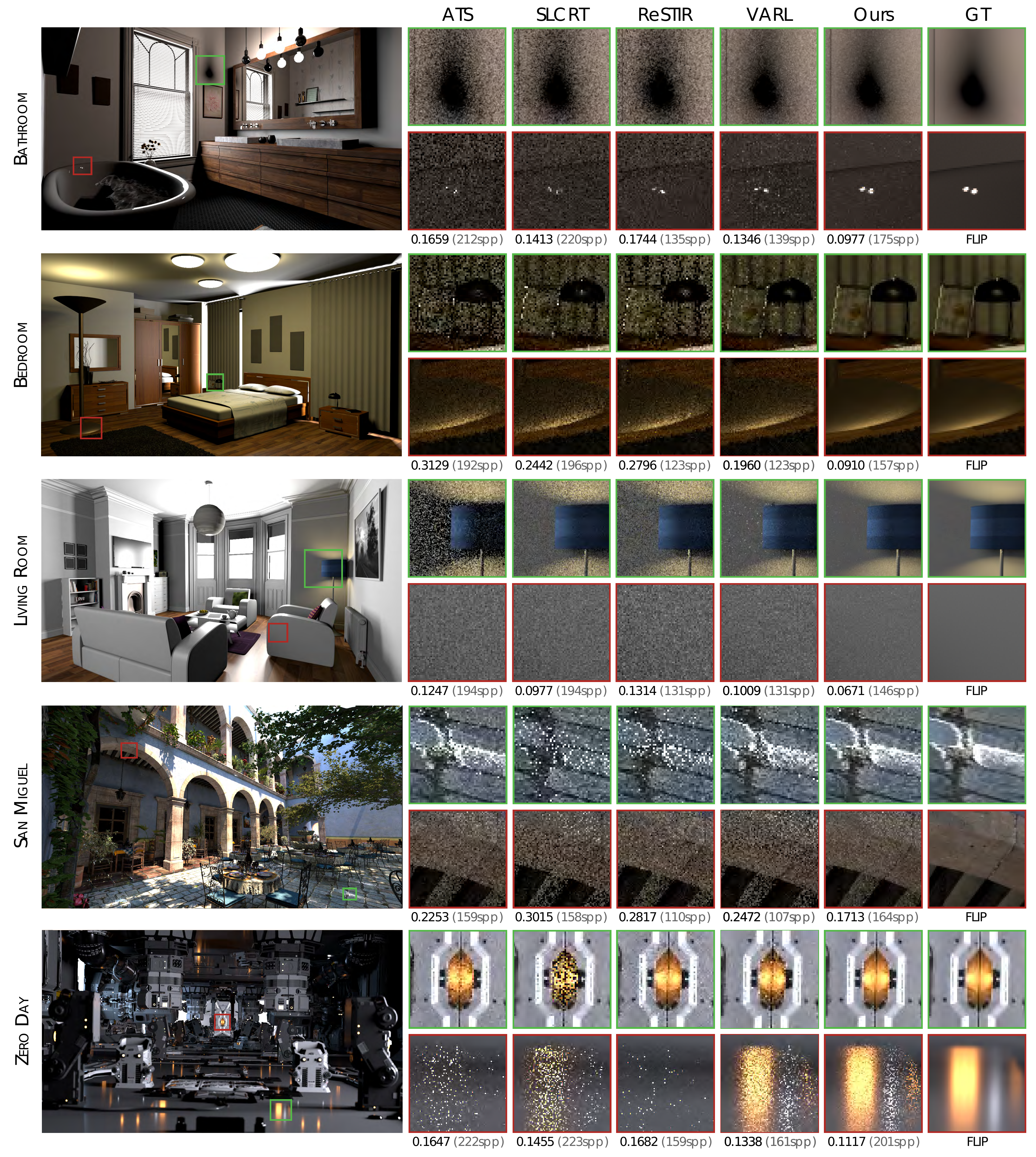}
\caption{\rev{Equal-time comparison of ATS~\cite{Estevez_2018_CGIT}, SLCRT~\cite{lin2020real}, ReSTIR~\cite{Bitterli_2020_TOG}, VARL~\cite{Wang_2021_TOG}, and our method. The time budget increases with scene complexity and resolution. For \textsc{Bathroom}, \textsc{Bedroom} and \textsc{Living Room}, the budget is 5 seconds; \textsc{Zero Day} and \textsc{San Miguel} 10 seconds.}}
\label{fig:equaltime}
\end{figure*}

\begin{figure*}
\includegraphics[width=\textwidth]{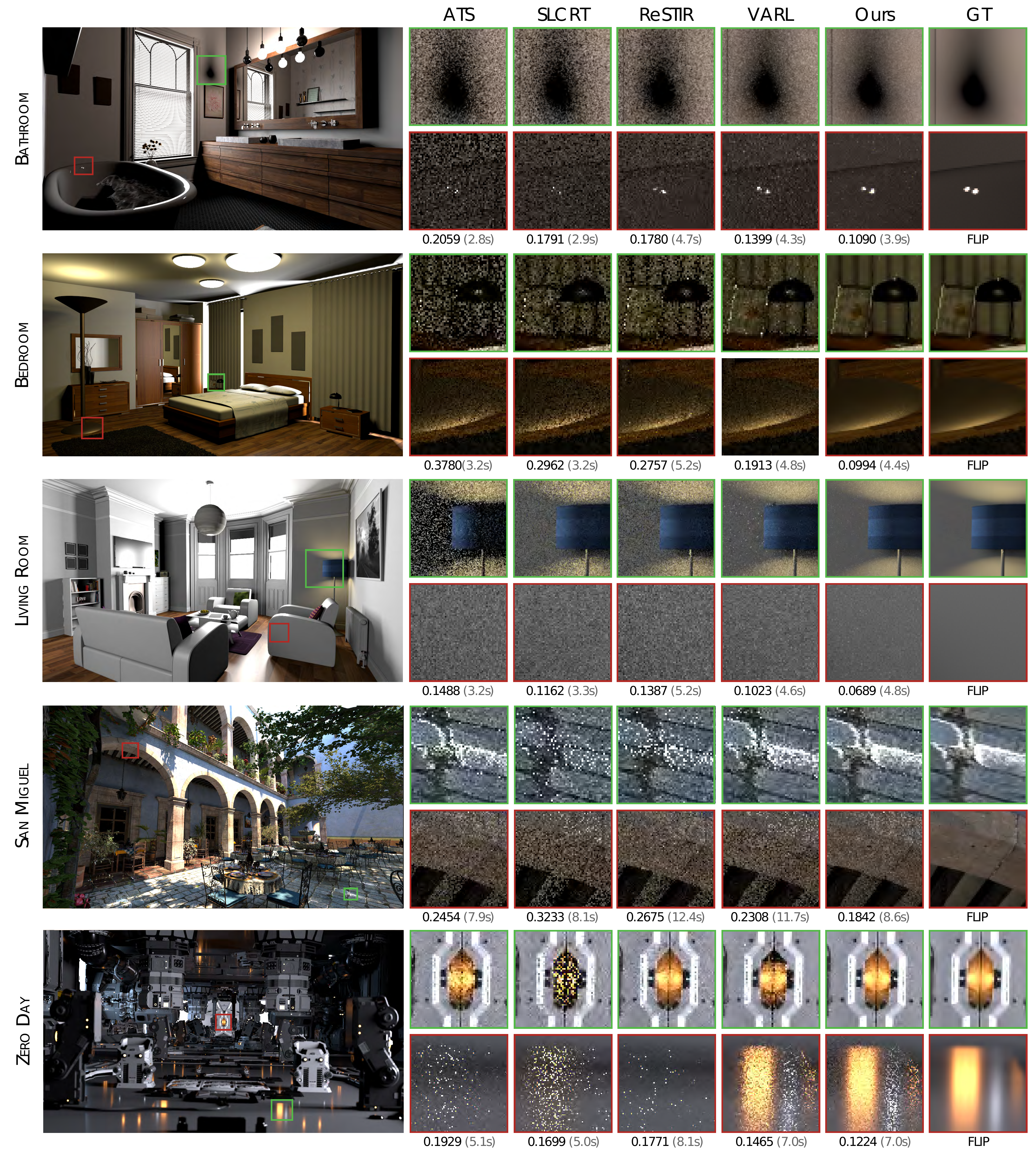}
\caption{Equal-sample comparison against several state-of-the-art methods. We use 128 spp for all the scenes.}
\label{fig:equalspp}
\end{figure*}

\begin{figure*}
\centering
\includegraphics[width=\textwidth]{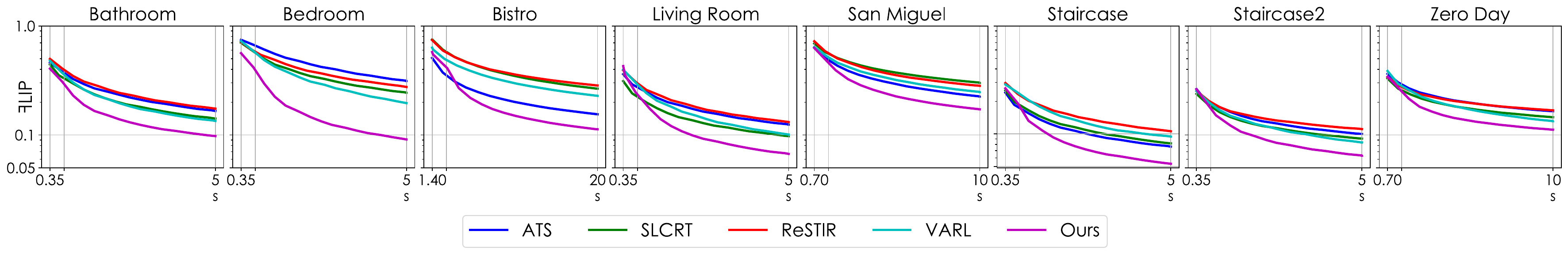}
\vspace{-0.2in}
\caption{\rev{Equal-time convergence plots of all the approaches on the eight scenes. The time budget increases with scene complexity. For all scenes, the second vertical line is approximately where our method stops learning and uses the learned distributions to sample the remaining paths.}}
\label{fig:timeconvergenceplots}
\end{figure*}

\begin{figure*}
\centering
\includegraphics[width=\textwidth]{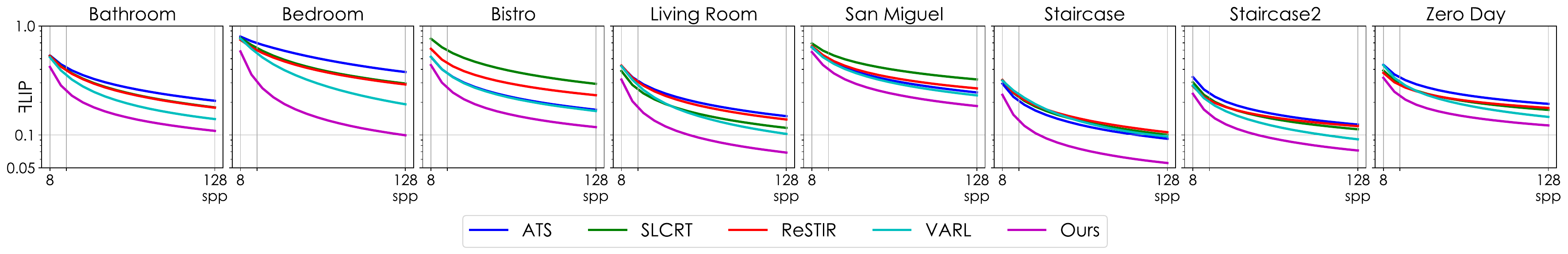}
\vspace{-0.2in}
\caption{Equal-sample convergence plots of all the approaches on the eight scenes from 8 to 128 spp. The second vertical line is approximately where our method stops learning and uses the learned distributions to sample the remaining paths.}
\label{fig:sppconvergenceplots}
\end{figure*}

\paragraph{\rev{Equal Time}} \revsp{All approaches are compared both quantitatively (Table~\ref{tab:equal_time}) and qualitatively (Fig.~\ref{fig:equaltime}) in an equal time setting with the time budget increasing with scene complexity and resolution. }\rev{For \textsc{Bathroom}, \textsc{Bedroom}, \textsc{Living Room}, \textsc{Staircase}, and \textsc{Staircase2}, we allocate a budget of 5 seconds to each approach. For \textsc{San Miguel} and \textsc{Zero Day}, we allocate 10 seconds, while, for \textsc{Bistro}, 20 seconds are allocated. As shown, our approach produces significantly better results than the others, both numerically and visually. In particular, our method best handles specular reflections (see the red insets for \textsc{Bathroom} and \textsc{Zero Day} in Fig.~\ref{fig:equaltime}), and seamlessly learns local lighting variations such as shadows (see the green insets for \textsc{Bathroom} and \textsc{San Miguel}).}


\paragraph{\rev{Equal Sample}} \rev{We compare all the approaches in an equal sample setting to evaluate their effectiveness regardless of speed. We render each scene using 128 samples per pixel, report the \reflectbox{F}LIP~\cite{Andersson_2020} metric in Table~\ref{tab:equal_spp}, and compare all methods qualitatively in Fig.~\ref{fig:equalspp}. Similar to equal-time results, our method ranks first in all scenes despite training for only 15\% of the allocated samples. Additionally, our approach shows competitive runtime against fast tree-based approaches, highlighting the small overhead of our neural training and sampling strategies.
}

\paragraph{\rev{Convergence Analysis}} \rev{To understand the convergence behavior of our approach, we plot the \reflectbox{F}LIP~\cite{Andersson_2020} metrics of rendered images in equal-time (Fig.~\ref{fig:timeconvergenceplots}) and equal-sample (Fig.~\ref{fig:sppconvergenceplots}). Overall, our approach consistently produces better results than all other methods, especially at higher spp counts. We note that, in some scenes, other methods produce better results initially in equal-time scenarios, but our performance improves significantly once the training phase ends and we can trace more samples.}

\subsection{\rev{Ablations}}
\begin{figure}[t]
\centering
\includegraphics[width=1.0\linewidth]{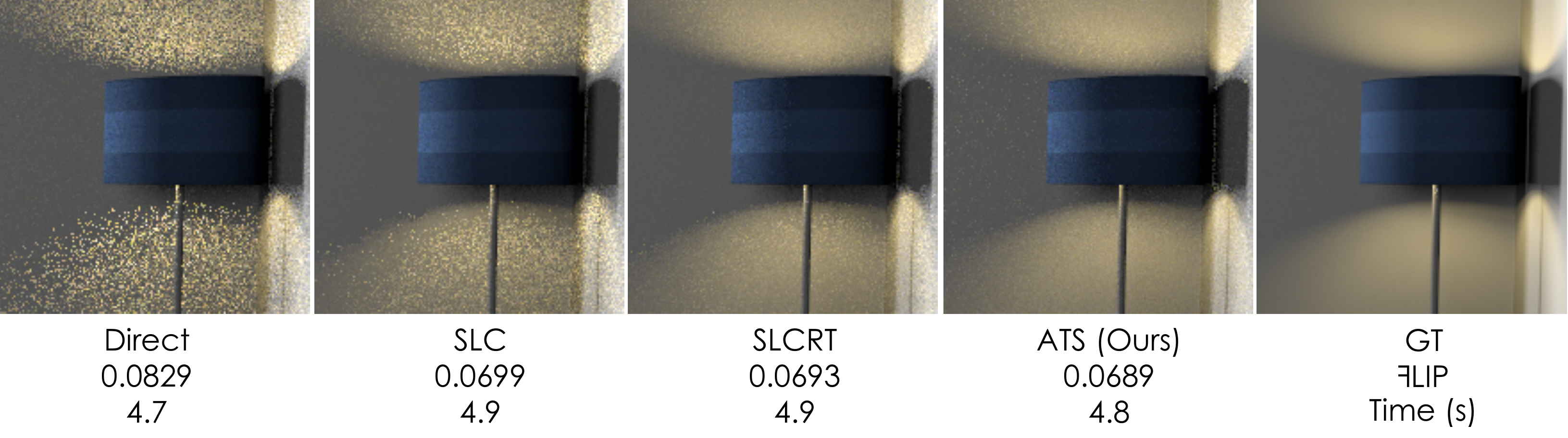}
\vspace{-0.2in}
\caption{\revsp{Comparison of residual learning strategy }\rev{using SLC~\cite{Yuksel_2019_HPG}, SLCRT~\cite{lin2020real}, and ATS~\cite{Estevez_2018_CGIT} baseline methods} against the version of our \rev{technique} where the cluster probabilities are directly estimated by the network (Direct). As shown, residual learning \rev{with any baseline method} produces significantly better results \rev{than direct estimation} with minimal timing overhead.}
\vspace{-0.1in}
\label{fig:residual}
\end{figure}

\paragraph{\rev{Residual Learning}} \rev{Figure~\ref{fig:residual} compares our proposed residual learning approach using SLC~\cite{Yuksel_2019_HPG}, SLCRT~\cite{lin2020real}, and ATS~\cite{Estevez_2018_CGIT} baseline methods to direct network prediction. The results show that residual learning greatly speeds up convergence while adding only minimal runtime overhead, with ATS chosen as our baseline since it ranks first.}

\begin{figure}[t]
\centering
\includegraphics[width=1.0\linewidth]{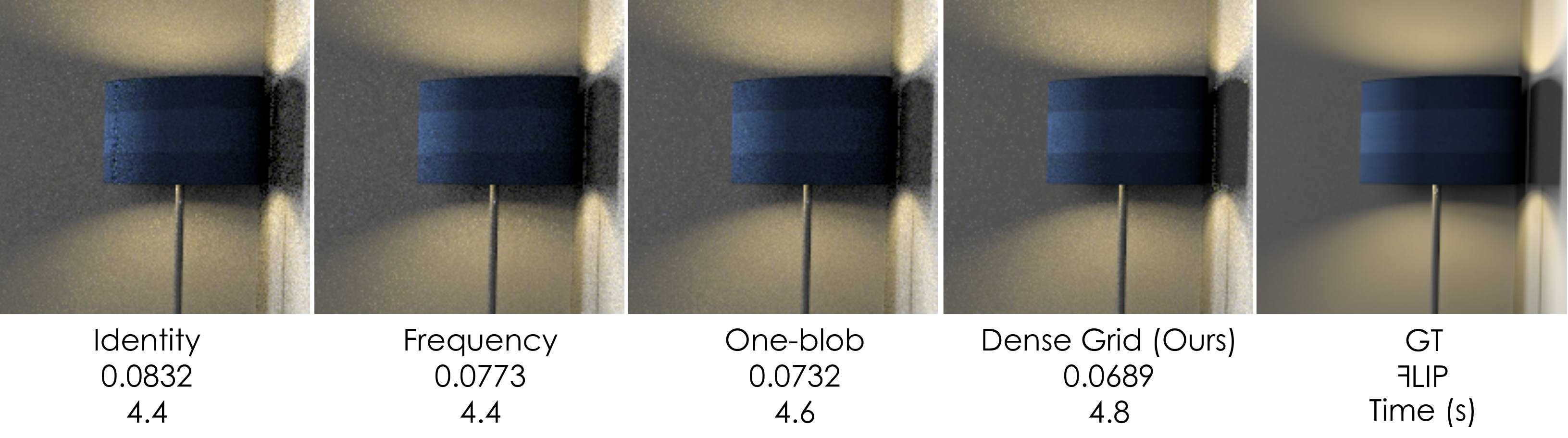}
\vspace{-0.2in}
\caption{\revsp{Evaluation of }\rev{input encodings for the position in the equal-sample \textsc{Living Room} scene. }\revsp{As shown, }\rev{the learnable dense grid~\cite{Dong_2023} performs best with a small overhead.}}
\vspace{-0.1in}
\label{fig:inputencoding}
\end{figure}

\paragraph{\rev{Input Encoding}} \revsp{Encodings for input position are evaluated in an equal-sample setting in Fig.~\ref{fig:inputencoding}. It compares identity, frequency~\cite{Mildenhall_2020} (12 frequencies), one-blob~\cite{Muller_TOG_2019} (32 bins), and the learnable dense grid~\cite{Dong_2023} encodings, with the latter being chosen because it exhibits the least noise.}

\begin{figure}[t]
\centering
\includegraphics[width=1.0\linewidth]{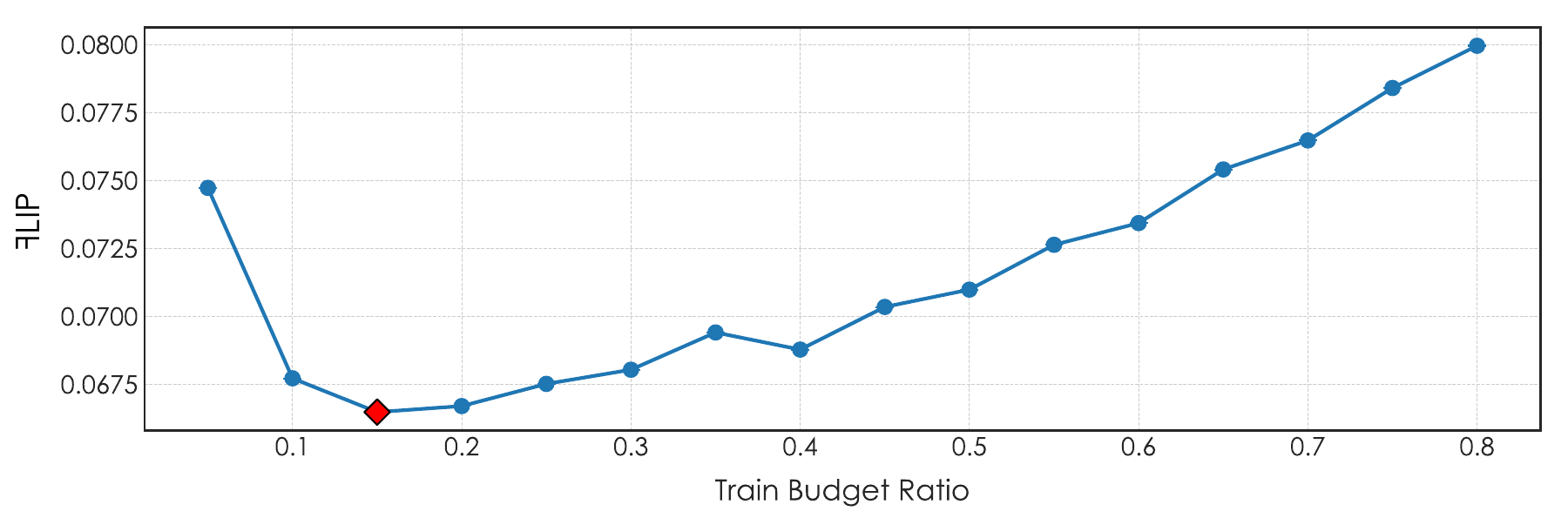}
\vspace{-0.25in}
\caption{\revsp{Impact of} \rev{training budget ratios on the quality of the final render, measured in \reflectbox{F}LIP~\cite{Andersson_2020}, for the \textsc{Living Room} scene with 5 seconds of time budget. We choose the optimal ratio of 15\%.}}
\vspace{-0.05in}
\label{fig:trainingbudget}
\end{figure}

\paragraph{\rev{Training Budget Ratio}} \revsp{Figure~\ref{fig:trainingbudget} shows that training budget ratios significantly influence quality in an equal-time setting. A 15\% training ratio achieves the highest quality in the 5-second \textsc{Living Room} scene, and similar trends are observed across all scenes.}

\paragraph{\rev{Effect of Continuous Representation}} \revsp{Table~\ref{tab:continuous_and_cluster} (left) evaluates the impact of our continuous representation versus a discrete alternative. }\rev{The discrete version takes as input the center of the nearest uniform grid cell with resolution 32x32x32 and the normalized outgoing ray direction discretized into 8 buckets per dimension~\cite{Wang_2021_TOG}. As seen, our method benefits from continuous input in all scenes, demonstrating its ability to encode local variations.}

\begin{table}[t]
\setlength\tabcolsep{4pt} 
\centering
\caption{\revsp{Analysis of }\rev{the effect of continuous representation and the cluster level. On the left, }\revsp{the results }\rev{show that our approach is superior to a discrete version in all scenes. On the right, 
 }\revsp{the comparisons indicate }\rev{that the cluster level $k=6$ performs best on most scenes. We use the \reflectbox{F}LIP~\cite{Andersson_2020} metric and highlight the lowest (best) scores in \textbf{bold}.}}
\vspace{-0.05in}
\label{tab:continuous_and_cluster}
\footnotesize
\begin{tabular}{lcc|ccc}
\toprule
 & \multicolumn{2}{c}{Effect of Continuous Representation} & \multicolumn{3}{c}{Cluster Level} \\
\cmidrule(lr){2-3} \cmidrule(lr){4-6}
      & Ours Discrete & Ours & k=4 & k=6 & k=8 \\
\midrule
\textsc{Bathroom}    & 0.1138      & \textbf{0.1090}      & \textbf{0.0969} & 0.0977 & 0.1081 \\
\textsc{Bedroom}     & 0.1027      & \textbf{0.0994}      & 0.0924        & \textbf{0.0910} & 0.1021 \\
\textsc{Bistro}      & 0.1321      & \textbf{0.1217}      & 0.1269        & \textbf{0.1157} & 0.1237 \\
\textsc{Living Room} & 0.0778      & \textbf{0.0689}      & \textbf{0.0617} & 0.0665 & 0.0668 \\
\textsc{San Miguel}  & 0.1891      & \textbf{0.1841}      & 0.1856        & \textbf{0.1713} & 0.1782 \\
\textsc{Staircase}   & 0.0732      & \textbf{0.0695}      & 0.0683        & \textbf{0.0630} & 0.0672 \\
\textsc{Staircase2}  & 0.0758      & \textbf{0.0721}      & 0.0712        & \textbf{0.0646} & 0.0673 \\
\textsc{Zero Day}    & 0.1387      & \textbf{0.1224}      & 0.1230        & \textbf{0.1117} & 0.1377 \\
\bottomrule
\end{tabular}
\vspace{-0.1in}
\end{table}

\paragraph{\rev{Cluster Level}} \rev{We assess how varying cluster levels (see Fig.~\ref{fig:light_hierarchy}) affects the final result. Table~\ref{tab:continuous_and_cluster} (right) shows numerical comparisons for three cluster levels ($k=4,6,8$) on all scenes. }\revsp{The comparisons indicate that the chosen $k=6$ performs best in most scenes.}
\section{Conclusion, Limitations, and Future Work}

We have presented a neural light importance sampling approach for rendering scenes with many lights. Our method employs a neural network to predict spatially varying light selection distributions using local information at each shading point. The network is trained online by minimizing the KL-divergence between the learned and target distributions. To efficiently handle scenes with a large number of lights, we integrate our neural approach with existing light hierarchy techniques. Additionally, we introduce a residual training strategy to improve convergence. Extensive experiments demonstrate that our approach outperforms existing methods on challenging scenes with complex lighting.

\begin{figure}[t]
\centering
\includegraphics[width=1.0\linewidth]{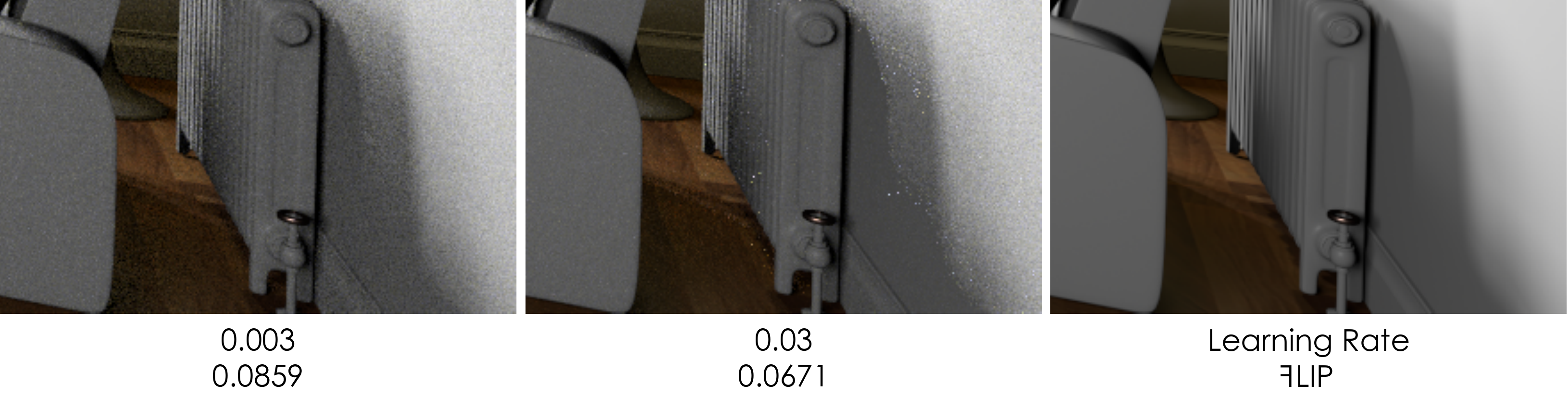}
\vspace{-0.2in}
\caption{\revsp{Comparison of }\rev{learning rates in the equal-time \textsc{Living Room} scene. Using an aggressive learning rate of $3\times10^{-2}$ yields lower noise levels than a more stable learning rate of $3\times10^{-3}$ at the cost of outlier fireflies.}}
\vspace{-0.2in}
\label{fig:lrlimitation}
\end{figure}

Despite its superior performance, our approach has some limitations. For instance, in some cases, \rev{it produces outlier fireflies due to our aggressive optimization strategy. During training, the network may assign low probability (in initial iterations) to important, unobserved light clusters. If a light with high power from such a cluster is sampled, it significantly contributes to the final pixel color, causing the outlier. Using a more conservative optimization strategy produces fewer outliers but results in slower convergence (Fig.~\ref{fig:lrlimitation}).}

Furthermore, we use a fixed global cut for the entire scene and rely on existing methods to sample lights within a cluster. Consequently, if the baseline method produces poor light sampling probabilities, our approach cannot fully correct these errors. One way to address this limitation is to locally adapt the cut, as demonstrated in existing methods~\cite{Wang_2021_TOG, Pantaleoni_2020}. \rev{We leave the combination of our approach with this strategy for future work}.

Moreover, we have focused solely on learning the light sampling distributions. However, sampling points within light sources also significantly impacts the variance of the Monte Carlo (MC) estimator, particularly for larger light sources. Learning the sampling distribution within a light source is another avenue for future work. Finally, we are interested in combining our approach with Zhu et al.'s method~\shortcite{Zhu_2021_ToG} to effectively handle scenes with a large number of complex luminaires. The key challenge lies in formulating a framework that seamlessly optimizes both our network and the luminaire network, which we leave for future exploration.

\begin{acks}
This project was funded in part by the NSF CAREER Award $\#2238193$. We thank Ryusuke Villemin for the valuable discussions, Vincent Serritella for generating the experimental test data, and Magnus Wrenninge and the Aurora simulation team for support. We are grateful to Wang et al.~\shortcite{Wang_2021_TOG} for releasing the source code of their work. We would like to thank the following artists for sharing their scenes and models that appear in our figures: Mareck (\textsc{Bathroom}), SlykDrako (\textsc{Bedroom}), Amazon Lumberyard~\shortcite{ORCAAmazonBistro} (\textsc{Bistro}), Jay-Artist (\textsc{Living Room}), Guillermo M. Leal Llaguno (\textsc{San Miguel}), Wig42 (\textsc{Staircase}), NewSee2l035 (\textsc{Staircase2}), and Mike Winkelmann~\shortcite{ZeroDay} (\textsc{Zero Day}).
\end{acks}

{\small
\bibliographystyle{ACM-Reference-Format}
\bibliography{bibliography}
}

\include{Sections/7_figures}

\end{document}